\documentclass[preprint,12pt]{aastex}
\usepackage{graphicx}

\shorttitle{Large Arc of Ionized Hydrogen}
\shortauthors{Reynolds, Sterling, \& Haffner}
  
\begin{document}

\title{Detection of a Large Arc of Ionized Hydrogen Far Above the Cas OB6
Association: A Superbubble Blowout into the Galactic Halo?}

\author{R. J. Reynolds} \author{N. C. Sterling} \author{L. M. Haffner}
\affil{Department of Astronomy, University of Wisconsin--Madison, 475
  North Charter Street, Madison, WI 53706\\
  Electronic Mail: {\tt reynolds@astro.wisc.edu, 
  sterling@uwalumni.com, haffner@astro.wisc.edu}}

\begin{abstract}
  
  The Wisconsin H$\alpha$ Mapper (WHAM) Northern Sky Survey has
  revealed a loop of H~II reaching 1300 pc from the Galactic midplane
  above the Cas OB6 association in the Perseus sprial arm.  This
  enormous feature surrounds and extends far above the ``W4 Chimney''
  identified by Normandeau et al.\ and appears to be associated with
  the star formation activity near the W3/W4/W5 H~II region complex.
  The existence of this ionized structure suggests that past episodes
  of massive star formation have cleared the H~I from an enormous
  volume above the Perseus arm, allowing Lyman continuum photons from
  O stars near the Galactic midplane to reach into the halo.

\end{abstract}


\section{Introduction}
\label{sec:intro}

The existence of the widespread, warm (10$^4$ K) ionized phase of the
interstellar medium is thought to be linked closely to star formation
activity (e.g., Zurita, Rosas, and Beckman 2000; Mathis 2000).  In
particular, it appears that hot massive stars, confined primarily to
widely separated stellar associations near the Galactic midplane, are
somehow able to photoionize a significant fraction of the gas not only in
the disk but also within the halo, 1-2 kpc above the midplane; no other
source with sufficient ionizing power has been identified (Reynolds 1993).  
The nature of this disk-halo connection is not understood.  For example,
the need to have a large fraction of the Lyman continuum photons from O
stars travel hundreds of parsecs through the disk seems to conflict with
the traditional picture of H~I permeating much of the interstellar volume 
near the Galactic plane.

It has been suggested that ``superbubbles'' of hot gas, especially
superbubbles that blow out of the disk (``galactic chimneys''), may sweep
large regions of the disk clear of H~I, allowing ionizing photons from the
O stars within them to travel unimpeded across these cavities and into the
halo (e.g., Norman 1991; Norman and Ikeuchi 1989).  Another possibility is
that the Lyman continuum radiation itself is able to carve out extensive
regions of H~II through low density portions of the H~I (e.g., Miller and
Cox 1993), perhaps creating photoionized pathways or ``warm H~II
chimneys'' that extend far above the midplane (Dove and Shull 1994; Dove,
Shull, and Ferrara 2000).  Although the existence of superbubbles has long
been established (e.g., Heiles 1984), direct observational evidence that
such cavities are actually responsible for the transport of hot gas and
ionizing radiation up into the Galactic halo is very limited.

The Wisconsin H$\alpha$ Mapper (WHAM) Northern Sky Survey has provided
strong evidence for superbubble blowout and for a disk-halo connection
that is allowing the transport of Lyman continuum photons from the
midplane into the Galactic halo above the ``W4 Chimney'' in the Perseus
spiral arm.  These results are presented below.

\section{Observations and Results}
\label{sec:obs}

WHAM is a high-throughput Fabry-Perot observing facility dedicated to the
detection and study of faint optical emission lines from the diffuse
ionized gas in the disk and halo of the Milky Way (Haffner 1999; Tufte
1997;  Reynolds, Haffner, and Tufte 1999).  It provides a 1$\degr$
diameter beam on the sky and produces a 12 km s$^{-1}$ resolution spectrum
over a 200 km s$^{-1}$ wide spectral window. This facility is located at
Kitt Peak, Arizona, and operated remotely from Madison, Wisconsin.  
Between 1997 January and 1998 September, WHAM carried out one of its
primary missions, a northern sky H$\alpha$ survey, consisting of
37,300 spectra on a $0\fdg85 \times 0\fdg98$ grid above $-30\degr$
declination. Each spectrum was centered near the LSR velocity and had a
sensitivity limit of approximately 0.1 R, where 1 R (Rayleigh) is
$10^6/4\pi$ photons cm$^{-2}$ s$^{-1}$ sr$^{-1}$ or $2.41 \times 10^{-7}$
erg cm$^{-2}$ s$^{-1}$ sr$^{-1}$ at H$\alpha$.  This H$\alpha$ survey
provides for the first time a radial velocity resolved map of the diffuse
ionized hydrogen over the sky comparable to earlier 21 cm surveys of the
neutral hydrogen.

Using the Dominion Radio Astrophysical Observatory (DRAO) high angular
resolution 21 cm observations, Normandeau, Taylor, and Dewdney (1996,
1997) found an H~I cavity surrounding the W4 H~II region and its
associated star cluster (OCl 352) in the Cas OB6 association, 2.2 kpc
distant in the Perseus arm.  Based on its upwardly directed conical
appearance in the H~I channel maps, Normandeau et al proposed that this
cavity is a Galactic chimney extending at least 4$\degr$ (150 pc) above
the midplane, the limit of the DRAO survey.  An H$\alpha$ image of the
region obtained by Dennison, Topasna, and Simonetti (1997) revealed an
H~II shell associated with the photoionized inner wall of this cavity,
which they were able to follow to 7$\degr$ (260 pc) from the midplane.  
The new WHAM survey provides the opportunity to examine the distribution
and kinematics of the H~II in this region to much higher Galactic
latitudes.  These observations suggest that the W4 Chimney is the lower
end of an enormous cavity that extends at least 30$\degr$ (1300 pc) from
the midplane.

A portion of the WHAM Northern Sky Survey from Galactic longitude
120$\degr$ to 155$\degr$ and from the Galactic equator to +40$\degr$
latitude is shown in Figure 1.  The lower portion of this H$\alpha$ map
includes the regions of active star formation associated with the W3/W4/W5
H~II regions in the Perseus spiral arm (e.g., Carpenter, Heyer, and Snell
2000).  The radial velocity interval $-65$ km s$^{-1}$ to $-55$ km
s$^{-1}$ was chosen so that the emission associated with the 2-3
kpc-distant Perseus arm is clearly revealed in the figure, while the
emission from the often brighter foreground H~II, having radial velocities
nearer the LSR, is excluded.  A striking feature in Figure 1 is the large
H$\alpha$ loop, arching away from the Galactic plane at $l = 132\degr$ and
$l = 145\degr$ and reaching a maximum latitude of $+30\degr$ near $l =
140\degr$.  The low angular resolution of the survey ($1\degr$) and
the limited dynamic range in this display wash out the H$\alpha$ emission
associated with the ``W4 Chimney''. The small horseshoe shaped feature
centered near $l = 134\degr$, b = +3$\degr$ outlines the location and
extent of the H$\alpha$ emission associated with the Chimney as observed
by Dennison et al (1997).  The H$\alpha$ emission line profiles in this
region of the sky generally consist of two radial velocity components, one
centered near the LSR, associated with relatively local gas within a few
hundred parsecs of the sun, and the other centered near $-50$ to $-60$ km
s$^{-1}$, associated with the gas in the Perseus arm.  This is illustrated
in Figure 2, which shows H$\alpha$ spectra from the WHAM survey in a
sequence of six directions cutting across the large loop.  In this
sequence the H$\alpha$ emission associated with the loop can be clearly
identified as the radial velocity component near $-60$ km s$^{-1}$, which
is most prominant in the three spectra from $l = 133\fdg1, b = +24\fdg6$
to $l = 134\fdg5, b = +22\fdg9$, indicating that the loop is approximately
3$\degr$ (100 pc) thick at that location.

Tables 1 and 2 list the centroid velocity, width, and intensity of the
H$\alpha$ emission associated with the W4 Chimney and the large H$\alpha$
loop, respectively.  The values are obtained from Gaussian fits to the
``Perseus arm component'' in the WHAM survey spectra at representative
positions along the H$\alpha$ ridges of the respective loops.  Each of the
selected directions along the large H$\alpha$ loop is indicated by an
``$\times$'' in Figure 1.  The emission measure for each of these 
directions was calculated from the relation
\begin{equation}
\mathrm{EM} = 2.75 \, \mathrm{I}_{\alpha} \, \mathrm{T}_{4}^{0.9} \,
\mathrm{e}^{{\tau}_{\alpha}} \;\; \mathrm{cm}^{-6} \, \mathrm{pc},
\end{equation} \
where I$_{\alpha}$ is the H$\alpha$ intensity in Rayleighs, T$_4$ is the
temperature in units of $10^4$ K, and ${\tau}_{\alpha}$ is the optical
depth at H$\alpha$ due to interstellar extinction.  Because this
loop is located far from the midplane, behind most of the H~I, we
estimated the extinction from the neutral hydrogen column density N(H~I)
in each direction (Hartmann and Burton 1997), using the relationship
between N(H~I) and ${\tau}_{\alpha}$ given by Bohlin et al (1978) and
Mathis (1990).  The values for ${\tau}_{\alpha}$ are approximately 0.3 at
$b \ga 20\degr$ and rise to about 0.8 for the lower latitude directions.  
The resulting emission measures are listed in the last column of Table 2
for T$_4 = 0.8$ (Haffner, Reynolds, and Tufte 1999).

At a distance of 2.2 kpc (see below), the top of the loop extends nearly
1300 pc above the midplane, and the distance between its ``feet'' (at
11$\degr$ latitude) is 500 pc.  The derived density and mass of this
structure depend somewhat on its geometry, whether it is a loop or a shell
viewed in projection.  The H$\alpha$ spectra provide little evidence for a
complete shell. The weak emission near $-60$ km s$^{-1}$ interior to the
loop is not brighter than that seen outside the loop, and there is no sign
of line splitting characteristic of an expanding shell.  Therefore,
assuming a loop of cylindrical shape, with an extent along the line of
sight equal to its extent on the sky, we find that the rms density of the
H$^+$ within the loop varies from 0.1 cm$^{-3}$ in the high latitude,
diffuse portion near $l = 140\degr, b = +25\degr$ to about 0.2 cm$^{-3}$
in the brighter lower latitude portion, near $l \approx
132\degr-134\degr$.  This implies that the recombination time ranges from
about 0.5 to 1.0 million years, the loop's total mass is approximately $1
\times 10^5$ M$_{\odot}$ (above 10$\degr$ latitude), and its gravitational
potential energy is approximately $3 \times 10^{51}$ ergs.

\section{Discussion}
\label{sec:dis}

A close relationship between this H$\alpha$ loop and the W4 chimney is
suggested both by appearance and kinematics.  Unfortunately, at Galactic
latitudes below about $10\degr$, the surface brightness of the large loop
is less than that of the more diffuse ionized gas in the Perseus arm and
can no longer be identified as a distinct feature.  However, the
downwardly cupped geometry of the large loop is strongly suggestive of it
being the high latitude complement to the upwardly cupped W4 Chimney.
Also, the radial velocities of the two features appear to blend smoothly
together.  The observed velocities of the large loop range from $-63$ km
s$^{-1}$ to $-45$ km s$^{-1}$ (Table 2), while in the W4 Chimney the
H$\alpha$ velocities range from $-60$ km s$^{-1}$ to $-39$ km s$^{-1}$
(Table 1).  The uppermost portions of the W4 Chimney ($b = +5\degr$ to
$+7\degr$) have a mean radial velocity ($\approx -50$ km s$^{-1}$) that
matches that of the lower portions ($b < +15\degr$) of the large loop. If
this interpretation is correct, then the large extent in longitude of the
H$\alpha$ loop relative to that of the W4 Chimney could be a result of
``blowout'' into the lower density halo beginning at about
7$\degr$--10$\degr$ (300--400 pc) above the midplane (see, for example,
Tenorio-Tagle, R\'{o}\.{z}yczka, and Bodenheimer 1990).  It is also
possible that the broader extent of the H$\alpha$ loop is the result of it
being created by a more distributed source of energy, that is, not just
the W4 region, but the entire W3/W4/W5 complex, which extends nearly 200
pc from $l = 134\degr$ to $l = 139\degr$ at $b \approx +1\degr$.

For the W4 Chimney, the associated H~II appears to be a photoionized wall
lining the inside surface of the H~I cavity (Dennison et al 1997).  For
the large H$\alpha$ loop, there is no clearly associated outer H~I wall.  
However, along the low longitude side ($l = 132\degr - 134\degr, b =
+14\degr$ to $+20\degr$), where the H$\alpha$ velocity is about $-62$ km
s$^{-1}$, there is a corresponding H~I ``filament'' offset approximate
$2\degr$ to lower longitude (i.e., to the outside of the loop) on the
$-62.5$ to $-55$ km s$^{-1}$ velocity interval maps of the Dwingeloo 21 cm
survey (Hartmann 1994).  This suggests that, as in the W4 Chimney, the
H~II loop is photoionized by sources within it.  Because of the relatively
short recombination time ($\lesssim 10^6$ yrs), the H$\alpha$ intensity of
the loop is then a measure of the current Lyman continuum flux (e.g.,
Tufte, Reynolds, and Haffner 1998).  Ignoring any geometric projection
effects, we estimate that the one-sided flux of ionizing radiation
incident on the loop ranges from $2.4 \times 10^6$ photons cm$^{-2}$
s$^{-1}$ on the brighter, low longitude side, where the H~II appears to be
ionization bounded, to $0.8 \times 10^6$ photons cm$^{-2}$ s$^{-1}$ at the
top near $l = 140\degr, b = +25\degr$.  Since there is no detected H~I
counterpart in this fainter portion, the loop may be density bounded, and
therefore the derived value for the incident ionizing flux for that
portion must be considered a lower limit.  The H$\alpha$ luminosity of the
entire loop implies a total hydrogen ionization rate of $3 \times 10^{48}$
s$^{-1}$.

The origin of most of the ionizing radiation appears to be the OCl 352
star cluster ($l = 134\fdg8, b = +1\fdg0$), which contains the largest
concentration of O stars (9) within the W3/W4/W5 complex and also appears
to be the most porous to its stars' ionizing radiation (Normandeau et al
1997).  The distance to OCl 352 is approximately 2.2 kpc (see Gray et al
1999).  The ionization rate of the loop thus requires 1.3\% of the
cluster's Lyman continuum luminosity of $2.3 \times 10^{50}$ photons
s$^{-1}$ (Basu et al 1999;  Dennison et al 1997).  However, since the loop
subtends only about 0.4 sr as seen by the cluster, this would require that
approximately 40\% of this luminosity escape the cluster, if it were
escaping isotropically.  This is larger than the 15\% estimated by Basu et
al (1999), but consistent with the value reported by Terebey et al (2000).
It is interesting to note that the Lyman continuum flux required to ionize
the loop (i.e., $2.4 \times 10^6$ photons cm$^{-2}$ s$^{-1}$) would be
totally absorbed by an H~I cloud of column density $8 \times
10^{18}$/n$_{\mathrm{H}}$ cm$^{-2}$, where n$_{\mathrm{H}}$ is the density
of the cloud.  Thus the presence of this large ionized loop, 1300 pc from
the midplane, indicates the existence of an enormous region of the
Galactic disk that is remarkably free of H~I.  Since most of the loop
appears to be density bounded, the extent of this H~I-free region could be
even greater than that outlined by the H$\alpha$.

Dennison et al (1997) have argued that the W4 Chimney is significantly
older (age of 6-10 Myr) than the OCl 352 cluster (2 Myr), suggesting that
the W4 Chimney was formed by an earlier generation of stars. There has in
fact been speculation that this region once contained a Giant Molecular
Cloud that was disrupted by an earlier generation of OB stars (Carpenter 
et al 2000).  The H$\alpha$ loop in Figure 1 extends 1000 pc
beyond the W4 Chimney and thus must be even older.  Numerical simulations
of superbubble growth suggest that blowouts into the halo, reaching
heights of 1000 pc, may take 10-20 Myr or more (e.g., Tomisaka 1998; Mac
Low, McCray, and Norman 1989).  Therefore, this entire structure, the loop
revealed by WHAM as well as the smaller W4 Chimney below it, seem to be
the product not of a single event, but a sequence of star formation
episodes in the W3/W4/W5 region over a period of at least 10-20 million
years.  The latest episode, centered near W4, is reenergizing and ionizing
the older, preexisting cavity, making it fluoresce in H$\alpha$.

\section{Summary and Concluding Remarks}
\label{con}

The large H$\alpha$ loop above the Cas OB6 association in the Perseus arm
reveals the existence of a huge H~I-free cavity apparently created by
energetic events associated with past star formation activity near the W4
Chimney discovered by Normandeau et al (1996).  The cluster OCl 352,
containing nine O stars, has reionized the outer walls of this cavity,
revealing its enormous extent on the WHAM H$\alpha$ survey.  The
morphology of this entire structure appears similar to that predicted by
numerical simulations of expanding superbubbles that have reached
``blowout'' into the galactic halo.

If such H~I-free cavities are common, they could explain how Lyman
continuum photons from O stars near the midplane are able to ionize large
regions of the disk and halo.  However, these cavities would be difficult
to identify, since their H$\alpha$ intensity (the evidence that they are
in fact a conduit for ionizing radiation) depends upon the Lyman continuum
luminosity of the stars that are currently within them.  A cavity 1000 pc
or more in size containing one or two O stars would have an H$\alpha$
surface brightnesses of only $\sim 0.1$ R, which would be lost in the
brighter, more diffuse H$\alpha$ background.  Nevertheless, the presence
of other large ionized loops revealed by the WHAM survey (e.g., Haffner,
Reynolds, and Tufte 1998) does suggest that this cavity above the Perseus
arm is not unique.

\acknowledgments

This work, including the operation of WHAM, was supported by the National
Science Foundation through grant AST96 19424.

\newpage

\begin{deluxetable}{cccc}
\tablecolumns{4}
\tablewidth{0pc}
\tablecaption{H$\alpha$ Parameters for the W4 Superbubble}
\tablehead{
\colhead{Direction} & \colhead{V$_{\mathrm{LSR}}$} & \colhead{FWHM} &
\colhead{Intensity} \\
\colhead{$(l, b)$} & \colhead{(km s$^{-1}$)} & \colhead{(km s$^{-1}$)} & 
\colhead{(R)\tablenotemark{a}}}
\startdata
134.8, +0.9  & -45 $\pm$ 1 & 29 $\pm$ 3 & 277 $\pm$ 19 \\
135.3, +1.7  & -43 $\pm$ 1 & 29 $\pm$ 3 & 142 $\pm$ 10 \\
135.9, +2.6  & -41 $\pm$ 1 & 30 $\pm$ 3 & 43.3 $\pm$ 3.0 \\
136.5, +3.4  & -42 $\pm$ 1 & 29 $\pm$ 3 & 14.2 $\pm$ 1.0 \\
136.1, +4.2  & -41 $\pm$ 1 & 30 $\pm$ 3 & 10.6 $\pm$ 0.8 \\
135.8, +5.1  & -48 $\pm$ 3 & 36 $\pm$ 5 & 4.6 $\pm$ 0.8 \\
135.5, +5.9  & -59 $\pm$ 2 & 28 $\pm$ 3 & 2.4 $\pm$ 0.3 \\
133.2, +6.8  & -46 $\pm$ 3 & 38 $\pm$ 4 & 2.1 $\pm$ 0.4 \\
133.5, +5.9  & -47 $\pm$ 2 & 33 $\pm$ 3 & 3.1 $\pm$ 0.4 \\
132.8, +5.1  & -46 $\pm$ 1 & 33 $\pm$ 3 & 6.9 $\pm$ 0.5 \\
133.2, +4.2  & -41 $\pm$ 3 & 49 $\pm$ 5 & 4.4 $\pm$ 0.8 \\
132.5, +3.4  & -40 $\pm$ 1 & 34 $\pm$ 3 & 4.9 $\pm$ 0.6 \\
132.9, +2.6  & -39 $\pm$ 1 & 31 $\pm$ 3 & 14.0 $\pm$ 1.0 \\
134.3, +1.7  & -43 $\pm$ 1 & 28 $\pm$ 3 & 79.9 $\pm$ 5.6 
\\
\enddata
\tablenotetext{a}{1R = $10^{6}/4\pi$ photons cm$^{-2}$ s$^{-1}$ sr$^{-1}$,
or $2.41 \times 10^{-7}$ erg cm$^{-2}$ s$^{-1}$ sr$^{-1}$ at H$\alpha$.}   
\end{deluxetable}

\newpage

\begin{deluxetable}{ccccc}
\tablecolumns{5}
\tablewidth{0pc} 
\tablecaption{Parameters for the Large H$\alpha$ Loop}
\tablehead{
\colhead{Direction} & \colhead{V$_{\mathrm{LSR}}$} & \colhead{FWHM} & 
\colhead{Intensity} & \colhead{EM} \\
\colhead{$(l, b)$} & \colhead{(km s$^{-1}$)} & \colhead{(km s$^{-1}$)} & 
\colhead{(R)\tablenotemark{a}} & \colhead{(cm$^{-6}$ pc)}}
\startdata 
144.3, +11.0  & -48 $\pm$ 2 & 24\tablenotemark{b} & 0.72 $\pm$ 0.09 & 3.8 $\pm$ 0.5\\
146.1, +19.5  & -54 $\pm$ 4 & 38\tablenotemark{b} & 0.70 $\pm$ 0.12& 2.2 $\pm$ 0.4 \\
144.9, +22.1  & -56 $\pm$ 6 & 30 $\pm$ 5 & 0.47 $\pm$ 0.09 & 1.5 $\pm$ 0.3\\
142.4, +23.8  & -53 $\pm$ 5 & 26\tablenotemark{b} & 0.38 $\pm$ 0.14 & 1.2 $\pm$ 0.5\\
140.0, +25.5  & -58 $\pm$ 5 & 27 $\pm$ 3 & 0.38 $\pm$ 0.13& 1.1 $\pm$ 0.4 \\
137.4, +24.6  & -56 $\pm$ 5 & 37 $\pm$ 7 & 0.81 $\pm$ 0.17 & 2.5 $\pm$ 0.6\\
134.9, +23.8  & -51 $\pm$ 1 & 29 $\pm$ 5 & 1.27 $\pm$ 0.17& 3.7 $\pm$ 0.5 \\
133.2, +22.1  & -42 $\pm$ 1 & 31 $\pm$ 4 & 1.76 $\pm$ 0.16 & 5.3 $\pm$ 0.5\\
133.6, +19.5  & -63 $\pm$ 3 & 23\tablenotemark{b} & 0.74 $\pm$ 0.16& 2.4 $\pm$ 0.6 \\
133.8, +17.8  & -60 $\pm$ 2 & 25 $\pm$ 3 & 1.07 $\pm$ 0.16 & 3.6 $\pm$ 0.6\\
132.0, +14.4  & -62 $\pm$ 4 & 29\tablenotemark{b} & 0.77 $\pm$ 0.17 & 3.4 $\pm$ 0.8\\
132.2, +11.9  & -45 $\pm$ 3 & 31\tablenotemark{b}  & 0.93 $\pm$ 0.17 & 5.1 $\pm$ 1.0\\
\enddata
\tablenotetext{a}{1R = $10^{6}/4\pi$ photons cm$^{-2}$ s$^{-1}$ sr$^{-1}$, or 
$2.41 \times 10^{-7}$ erg cm$^{-2}$ s$^{-1}$ sr$^{-1}$ at H$\alpha$.}
\tablenotetext{b}{Due to blending, the width of this component is highly 
uncertain.  An average width has been fixed in order to determine 
V$_{\mathrm{LSR}}$ and the intensity.}
\end{deluxetable}

\newpage

\begin{figure}[htbp]
  \begin{center}
    \includegraphics{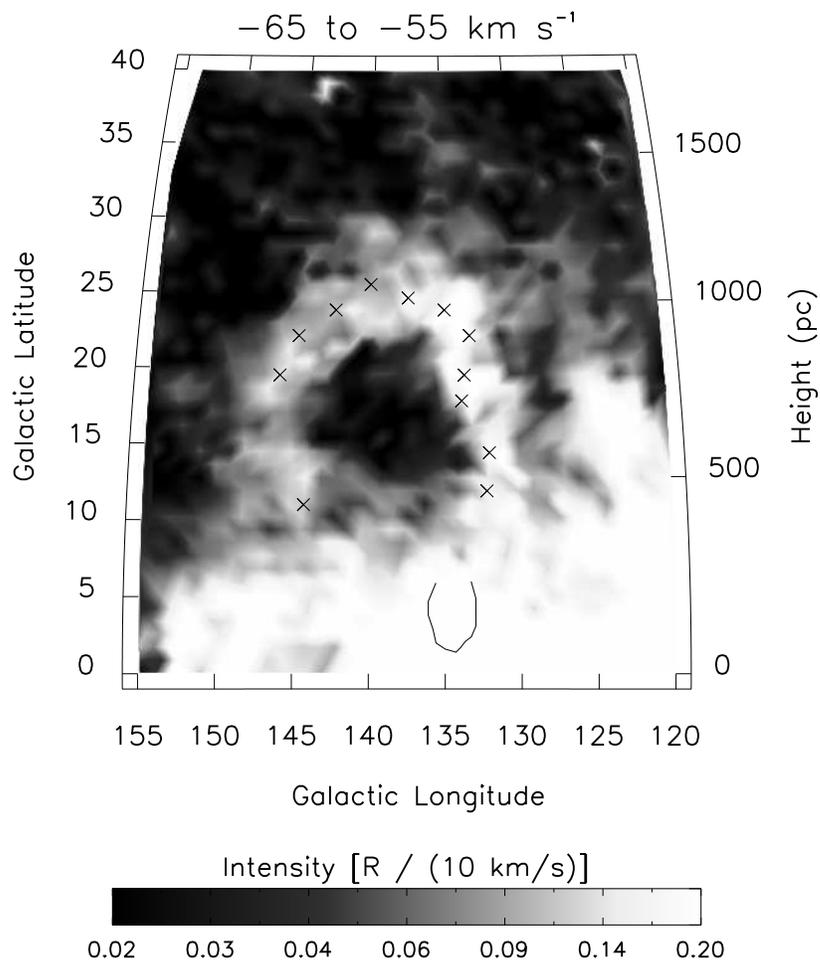}
    \vspace{-3.5in}
    \caption{A map of ionized interstellar hydrogen 
      above the Perseus spiral arm in the velocity interval $-65$ km
      s$^{-1}$ to $-55$ km s$^{-1}$ (from the WHAM Northern Sky
      Survey).  The horseshoe-shaped figure near $l = 135\degr, b =
      +3\degr$ indicates the location of the W4 Chimney identified by
      Normandeau et al (1996) and Dennison et al (1997).  Directions
      listed in Table 2 are denoted by an ``$\times$''.  The intensity
      scale is logarithmic. The vertical scale on the right side of
      the map denotes the height above the midplane for a distance of
      2.2 kpc.}
    \label{fig:loop}
  \end{center}
\end{figure}

\begin{figure}[htbp]
  \begin{center}
    \includegraphics{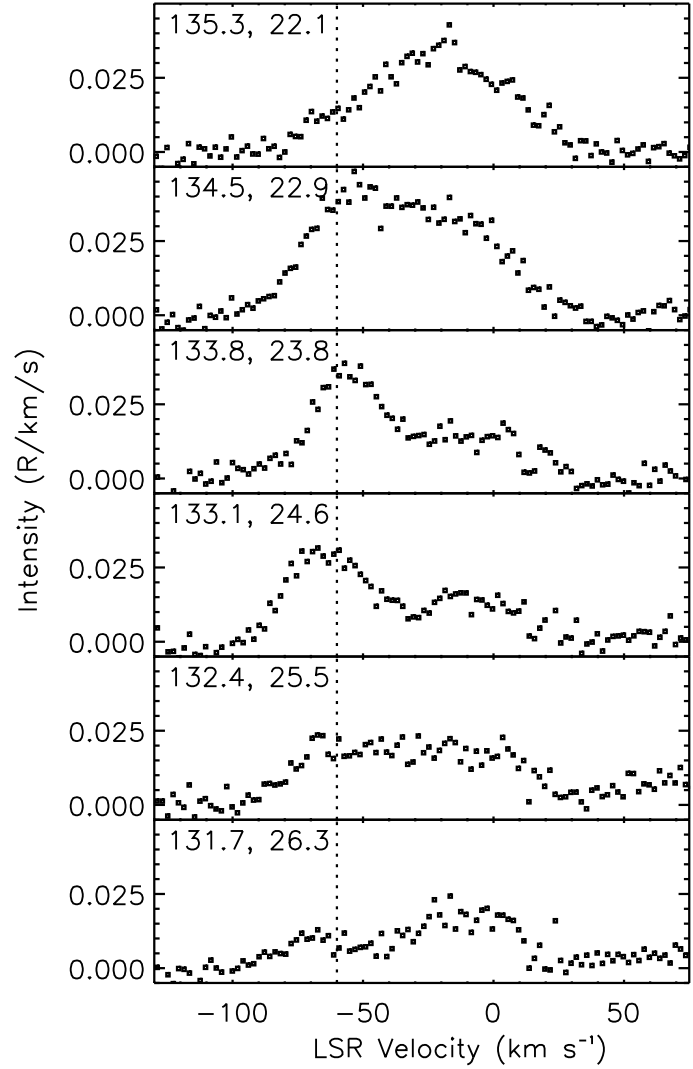}
    \vspace{-3.5in}
    \caption{Spectra from the WHAM Sky Survey 
      showing the H$\alpha$ line profiles at six positions that cut
      across the large H$\alpha$ loop near $l = 133\degr, b =
      +24\degr$.  Emission from the Perseus arm is associated with the
      velocity component centered near $-60$ km s$^{-1}$ (vertical
      dotted line).}
    \label{fig:spec}
  \end{center}
\end{figure}

\end{document}